\documentclass{aastex}

\usepackage{emulateapj5}
\input{epsf}

\usepackage{times}

\newcommand{\RXTE}{{\it RXTE}}

\newcommand{\lum}{\,{\rm erg}\,{\rm s}^{-1}}
\newcommand{\cminvsq}{\,{\rm cm}^{-2}}

\newcommand{\Ledd}{L_{\rm Edd}}
\newcommand{\Ldisk}{L_{\rm disk}}
\newcommand{\Lh}{L_{\rm hard}}
\newcommand{\Tin}{T_{\rm in}}
\newcommand{\Tbb}{T_{\rm bb}}
\newcommand{\Te}{T_{e}}

\newcommand{\Rin}{R_{\rm in}}
\newcommand{\Rc}{R_{\rm c}}
\newcommand{\Rg}{R_{\rm g}}
\newcommand{\taut}{\tau_{\rm T}}
\newcommand{\NH}{N_{\rm H}}
\newcommand{\grad}{{\rm o}}

\begin{document}

\slugcomment{The Astrophysical Journal Letters, in press}
 

\shorttitle{THERMAL COMPTONIZATION IN GRS 1915+105}
\shortauthors{VILHU ET AL.}

\title{Thermal Comptonization in GRS 1915+105}

\author{Osmi Vilhu,\altaffilmark{1} Juri Poutanen,\altaffilmark{2}
Petter Nikula,\altaffilmark{1} and Jukka Nevalainen\altaffilmark{3}
}

\altaffiltext{1}{Observatory, Box 14, FIN-00014 University of Helsinki, Finland;
osmi.vilhu@helsinki.fi; petter.nikula@helsinki.fi}
\altaffiltext{2}{Stockholm Observatory, SE-133 36 Saltsj\"obaden, Sweden; 
juri@astro.su.se}
\altaffiltext{3}{ESA/ESTEC Astrophysics Division, Box 299, 2200 AG Noordwijk,
The Netherlands; jnevalai@astro.estec.esa.nl}

\begin{abstract}
The Rossi X-ray Timing  Explorer  (\RXTE) data of GRS 1915+105 from several
observing  periods  are  modeled  with  a  thermal  Comptonization  model.
Best-fit  models  indicate that there is a strong  correlation  between the
inner  disk  temperature  and the disk  luminosity.  The  hard  Comptonized
luminosity  does not  depend  significantly  on the total  luminosity.  The
spectral  hardness of the Comptonized  radiation, the fraction of seed soft
photons scattered by the Comptonizing cloud, its Thomson optical depth, and
the fraction of the total power dissipated in the optically thin hot plasma,
all strongly  anticorrelate  with the  luminosity.  We find that the inner
disk radius is almost constant and that the hot Comptonizing corona shrinks
at high  luminosities.  We note that the fits using  
{\sc  xspec}  diskbb +
power law  model   underestimate  the amplitude  of the blackbody
component (and therefore the  corresponding  size of the emitting  region)
and overestimate  the  absorption  column  density  and  the  total,
corrected for absorption, luminosity.  
\end{abstract}


\section{Introduction}

The X-ray  transient GRS 1915+105 was  discovered by  Castro-Tirado  et al.
(1992) using the WATCH  all-sky  monitor on the {\it GRANAT}  satellite.  Since
then it has been one of the most  luminous  X-ray  sources in the sky.  The
{\it Rossi X-ray Timing Explorer} (\RXTE) has been monitoring it frequently
and a rich  pattern of  variability  has emerged  from these data with time
scales from years down to 15 msec (see e.g., Morgan, Remillard, \& Greiner
1997; Muno,  Morgan, \& Remillard  1999;  Belloni et al.  2000).  Often the
overall spectral shape in the 2--50 keV energy range has been modeled with
a disk blackbody  accompanied  by a power law tail  (e.g.  Belloni et al.
1997; Muno et al.  1999).

A power law,  however, is not a good  approximation  to the  Comptonization
spectrum  in the energy  range  close to the peak of the blackbody.  This
fact  inspired  us to use a  thermal  Comptonization  code by  Poutanen  \&
Svensson  (1996, PS96) to model the spectrum.  Vilhu \&  Nevalainen  (1998)
applied  a  similar  analysis  to a  selected  set of  observations  of GRS
1915+105.  A spherical  geometry for the hot  Comptonizing  plasma cloud is
assumed where the seed soft photons are coming from the  surrounding  cool disk
which has some overlap with the central  cloud  (Poutanen,  Krolik, \& Ryde
1997).  A geometry, with no overlap between the disk and the hot cloud
(``corona''), can correspond to a physical  situation when the central part
of the disk is overheated  (Beloborodov  1998) at large  accretion  rates.
The central hot cloud  can also be related to the innermost part of the jet.  In
the opposite  situation, when there is a large overlap between the disk and
corona, we arrive at a simple  disk-corona  model (e.g.  Haardt \& Maraschi
1993;  Svensson \&  Zdziarski  1994; Stern et al.  1995).  Physically  this
could  correspond  to the release of a large  fraction of the total energy in
the surface layers of the accretion disk, for example, due to  annihilation
of buoyant  magnetic  fields (e.g.  Tout \& Pringle  1992;  Miller \& Stone
2000).  The  adopted  geometry  is  thus  quite  generic  and it    can
represent well the X-ray emitting region in a number of physical situations.

An  important  question is whether  the energy  distribution  of  electrons
responsible for  Comptonization  is thermal or non-thermal.  In Cygnus X-1,
for  example,  when the  spectrum  is hard,  electrons  are mostly  thermal
(Gierli\'nski  et al.  1997;  Poutanen  1998),  while  in the  soft  state,
Comptonization  probably  proceeds in hybrid,  thermal/non-thermal  plasmas
(Poutanen  \& Coppi  1998;  Gierli\'nski  et al.  1999).  The lower  energy
photons (below $\sim 20$ keV) are produced  mostly by a thermal  population
of  electrons,  and the high  energy  tail  extending  to  $\sim  1$ MeV is
produced by single scattering off non-thermal electrons.  Up to date, there
are no detailed  spectral  studies of GRS 1915+105 using  physical  models.
However,  if one uses Cyg X-1 as an  analogy,  one can argue  that  thermal
Comptonization  probably  dominates  the spectra  below 20--50 keV.  In the
present paper, a thermal Comptonization model is applied to the \RXTE\ data
of  GRS  1915+105   collected  from  several  observing   intervals  during
1996--1997.

\section{Observations}

We collected 36  Proportional  Counter  Array (PCA) and  High-Energy  X-ray
Timing Experiment  (HEXTE)  observations of GRS 1915+105,  performed during
1996--97,  from the  `production'  archive of \RXTE, with typical  observing
times of a few hours.  The selection  procedure was rather random, its main
purpose  was to  extract  a  sufficient  number  of low,  medium  and  high
luminosity  states of the system.  Figure~\ref{asm}  shows the ASM (All Sky
Monitor) light curve with the selected PCA and HEXTE  observations  marked.
The data with 128 (PCA Standard 2 data) and 64 (HEXTE) channels of spectral
information  and 16 s  temporal  resolution  of  five  PCU's  (proportional
counter  units) and both HEXTE  clusters  were  used.  The  background  was
subtracted (using PCAbackest and HEXTE rocking)  although its effect is not
crucial, since we limited the PCA and HEXTE  spectra in the range 2--20 keV
and 15--60 keV, respectively.

Inside  each of the 36  observation  periods  the  data  were  binned  into
separate  luminosity groups (1--5) to accumulate a spectrum (using PCA count
rate criterion),  resulting in 101 individual spectra.  Single spectra were
used to  represent  the  non-variable  LULL-state  observations  (the  long
minimum phase in the middle of Fig.~\ref{asm}),  while strongly oscillating
observations  were  divided  into  5  luminosity   classes.  These  5-class
observations were 20402-01-33-00 (3000--26000  counts s$^{-1}$ , $\kappa$), 
20402-01-37-01 (3000-35000 counts s$^{-1}$, $\lambda$), 20402-01-43-00  
(5000--45000 counts s$^{-1}$, mixed $\beta$
and $\mu$) and 20402-01-44-00 (5000--40000  counts s$^{-1}$, $\beta$) 
where the ranges of
the five PCU's  count rates  and the  variability  types by  Belloni  et al.
(2000) are given in the parentheses.  On the average a typical  observation
was split into three classes.


\medskip
\centerline{\epsfxsize=8.8cm \epsfbox{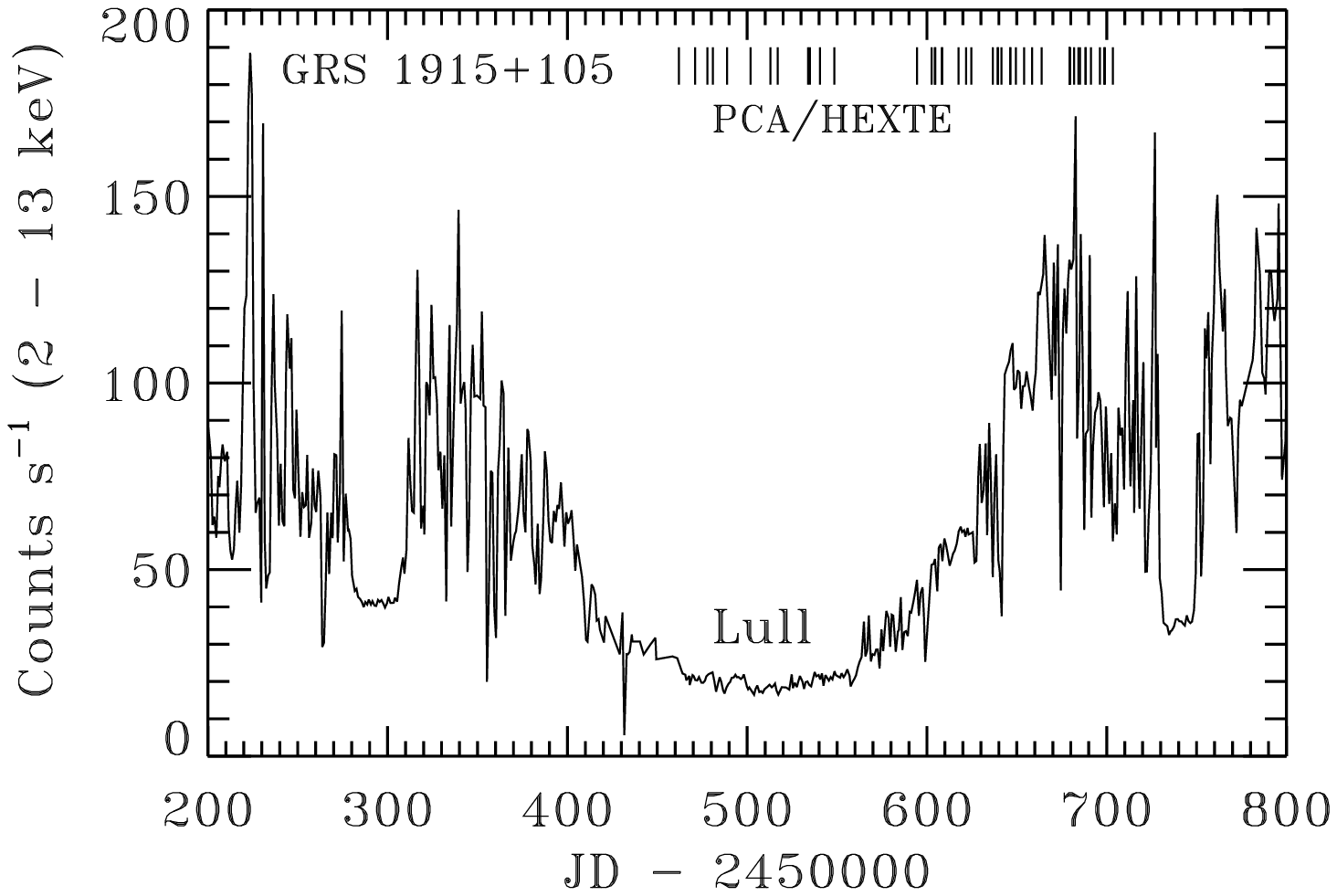}}
\figcaption{
The ASM light curve (2--13 keV) of GRS 1915+105 with the PCA/HEXTE 
observations used in this paper marked with short vertical lines.
\label{asm}}
\medskip

\section{The model}

The Comptonizing  cloud (corona) is represented as a homogeneous  sphere of
radius $\Rc$ and Thomson optical depth $\taut$.  The electrons in the cloud
are  assumed  to  have  thermal  distribution  of  temperature  $\Te$.  The
optically  thick  disk  supplying  seed  photons  for  Comptonization   and
penetrating  into the central cloud has an inner radius $\Rin < \Rc$.  
Radiative transfer in the corona (Comptonization) is handled by the code of
PS96 for a  hemispherical  geometry  of the  corona.  In order to model the
radiative transfer in a sphere, the boundary condition at the bottom of the
hemisphere   is   modified.  A   photon   crossing   the   bottom   can  be
mirror-reflected  instead  of  being  absorbed  by  the  disk  (or  Compton
reflected) with the probability  defined by the ratio $\Rin/\Rc$  (Poutanen
et al.  1997).  The  angle-dependent  Compton  reflection  from the disk of
neutral  material is  computed  using  Green's  function  of  Magdziarz  \&
Zdziarski (1995).

The radial  dependence  of the disk  temperature  is that of the  classical
viscous disk $\Tbb(r) = \Tin(r/\Rc)^{-3/4}$ at $R>\Rc$ and $\Tbb(r) = \Tin$
inside the  coronal  region  (between  $\Rin$ and  $\Rc$).  There are a few
reasons  why such a   temperature  profile  is  chosen.  First, if a
large  fraction of the total  power is  dissipated  in a corona which has
large  scale-height  and covers a significant  part of the inner cold disk,
the  reprocessing  of  coronal  hard  X-rays  can  produce  a  rather  flat
temperature  profile.  Second,  owing  to the  stress-free  inner  boundary
condition  at the  marginally  stable  orbit,  the  temperature  profile 
even  in a
standard  disk is rather flat between $3\Rg$ and $10\Rg$  
(where  $\Rg=2GM/c^2$).
Third,  if  $\Rin/\Rc  \rightarrow  0$, the  $r^{-3/4}$  profile  diverges.
Regarding  spectral  fitting, the actual  temperature  profile makes little
difference.  The local spectrum was assumed to be a blackbody.

The total  spectrum  contains a blackbody-type  component,  a  Comptonized
tail and a Compton reflection component.  To obtain both $\Te$ and $\taut$
is beyond the \RXTE\  capability,  therefore we fixed $k\Te$ at 70 keV (but
also made  numerous  fits with  $k\Te = 30$ keV and 150  keV).  Instead  of
$\taut$,  we  then  used  the  $y$-parameter  ($y  =  4\taut\Theta$,  where
$\Theta=k\Te/m_ec^2$)  which determines the spectral slope.  Thus, the free
parameters of the model are $\Rin/\Rc$,  $y$ and $\Tin$.  The  inclination
$i=70^{\grad}$  was used, assuming that the radio jet is  perpendicular  to
the disk (Mirabel \& Rodr{\'\i}guez 1994).  The important fact to notice is that
the relative normalization between the disk and the Comptonized  components
is not free, but is a function of the model parameters  mostly depending on
$\taut$ (or $y$) and $\Rin/\Rc$.

\section{Results}

The  PCA/HEXTE  spectra of GRS 1915+105  were fitted using {\sc xspec v.10}
(Arnaud   1996),  and   allowing   2\%   systematic   errors.  The  reduced
$\chi^2$-values  are less  than 2.2 in all data sets,  less  than 1.6 in 64
sets, and less than 1.2 in 32 sets (for 132  degrees of  freedom  using 138
energy  bins).  The neutral  hydrogen  column  density  $\NH$ was frozen at
$2.3\times10^{22}\cminvsq$.  This value is the best acceptable common value
for  spectra with high,  intermediate  and low  luminosities.  It is
lower than that used in many  previous  works (e.g.  Muno et al.  1999 used
$\NH=6.0  \times 10^{22}\cminvsq$).  However,  our  value is  close  to the
galactic  $\NH$ in the direction of GRS 1915+105  (using  Dickey \& Lockman
1990  data,  ftools/nh  gives  $\NH  =  1.75\times  10^{22}\cminvsq$, 
see also Rodr{\'\i}quez et al. 1994).  The
normalization factor between PCA and HEXTE data was allowed to vary freely,
with the best-fit always giving HEXTE/PCA = $0.70\pm 0.05$.

Figure~\ref{multi}  shows the  best-fit  results as a function  of the disk
luminosity  $\Ldisk$  (assuming  a  distance  of 12.5 kpc, see Chaty et al.
1996;  Fender et al.  1999)  which was  computed  from the  blackbody  flux
corrected for scattering in the corona and the $\cos i$ angular  dependence
of the disk  luminosity.  $\Ldisk$ and $\Tin$  together then  determine the
physical size of the system $\Rin$ and $\Rc$ (assuming blackbody emission).
The hard luminosity, $\Lh$, is computed from the (approximately  isotropic)
flux in the Comptonized component (plus reflection).

The inner disk temperature  $\Tin$ and $\Rin/\Rc$  strongly  correlate with
the disk luminosity  (Figs.~\ref{multi}a, d).  The coronal $y$-parameter
anticorrelates  with $\Ldisk$  depending on $\ln \Ldisk$  nearly
linearly   (Fig.~\ref{multi}b).  For  low  $\Ldisk$,  the  spectra  are
  hard, and for large  $\Ldisk$  the spectra  are soft.  The hard
Comptonized  luminosity, $\Lh$, is relatively constant and in some cases of
large  $\Ldisk$   significantly   decreases   (Fig.~\ref{multi}f).  The
geometry  of the  system  (determined  by  the  ratio  $\Rin/\Rc$)  changes
dramatically  when $\Ldisk$ varies.  For large $\Ldisk$, the  normalization
of the  Comptonized  component  decreases  relative to the disk  component,
leading to  $\Rin/\Rc\sim  1$.  In our model,
$\Rin/\Rc\rightarrow 1$ means that only a small fraction $\sim 3\%$ of soft
disk  photons are  getting  Comptonized,  while the  majority  reaches  the
observer  directly.  We would not advise to take the ratios  $\Rin/\Rc$  at
their face  value  since we made an  assumption  of a constant  $\Te$ (see
\S~\ref{sec:disc}).  One can  conclude,  however,  that the  number of soft
photons  that  are  intercepted  by  the  corona  (``scattered  fraction'')
decreases dramatically when $\Ldisk$ increases.

The inner disk radius (Fig.~\ref{multi}e) is rather constant (25--40 km,
without  relativistic and color corrections).  However, in two observations
(20402-01-43-00 and 20402-01-44-00)  the radii increased above 50 km during
their lowest luminosity phases ('downs' in  Fig.~\ref{multi}e,  while 'ups'
in  Fig.~\ref{multi}e  represent their high luminosity  phases).  These two
particular  observations  have  $\beta$-type  light curves (see Fig.~2X in
Belloni  et al.  2000)  where  the  long  minimum  phases  last  10--15  min
(mini-lulls).  It is  interesting  to  note  that  these  two  observations
preceded  by 10 days those by Mirabel  et al.  (1998)  with a similar  light
curve  during  which  ejections  of weak jets  were
discovered  (seen in IR and  radio).

To permit a  comparison  with models  using a power law, we also fitted all
our spectra  with the `wabs + diskbb + powerlaw'  model  of {\sc xspec}
  (which we call the
power law model  hereafter) fixing  $\NH=4.5\times 10^{22}
\cminvsq$.   The best-fit  power law photon index,  $\Gamma$, is
shown in  Fig.~\ref{multi}c.  One sees an anticorrelation  between $\Gamma$
and  $y$-parameter  which is well represented by the relation  $\Gamma=9/(4
y^{2/9})$ (Beloborodov 1999).

 
\medskip
\centerline{\epsfxsize=7.5cm \epsfbox{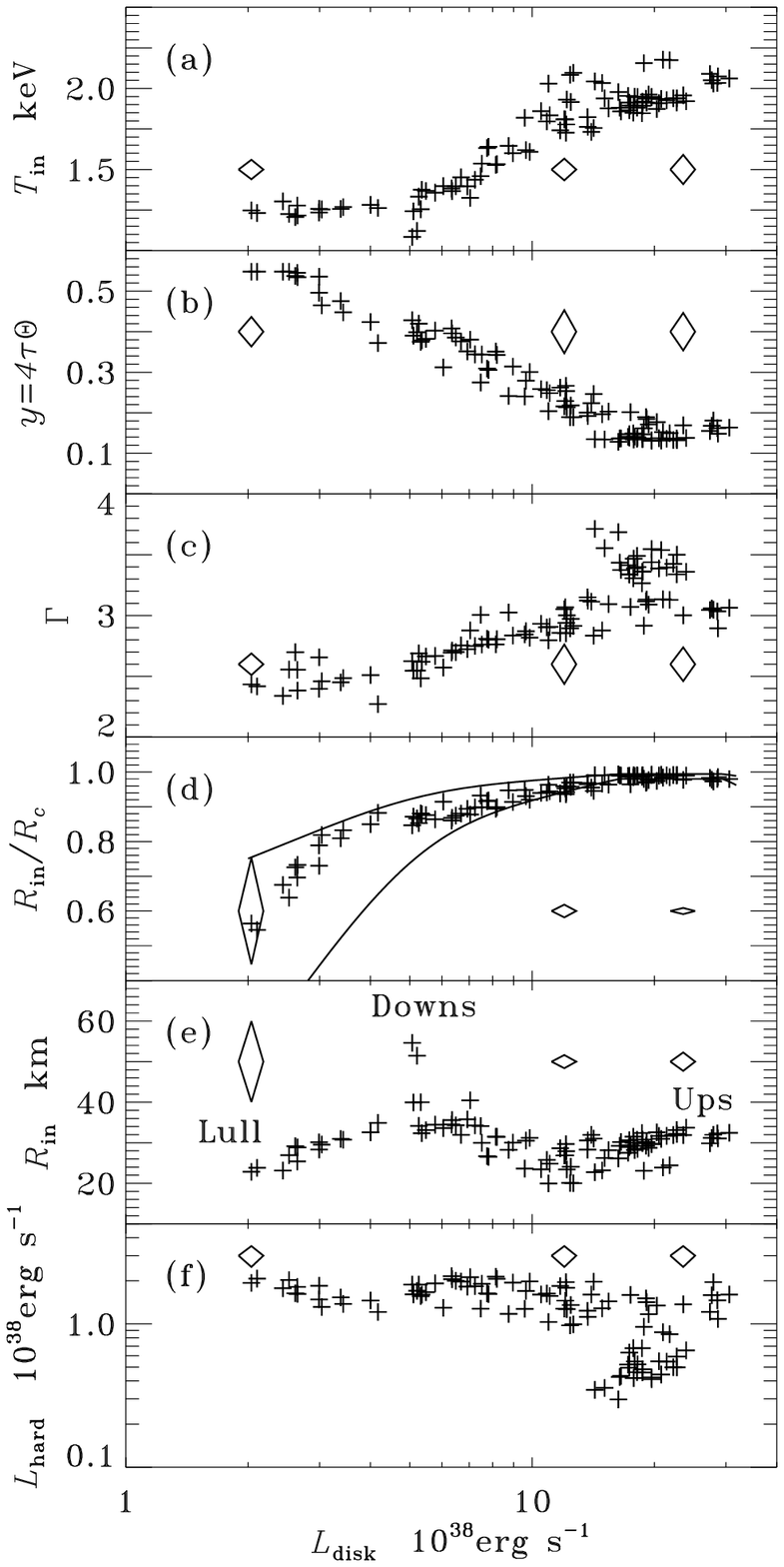}}
\figcaption{
The best-fit results as functions of the disk luminosity, $\Ldisk$.
The  typical $1\sigma$ error  bars are shown by diamonds.
The coronal temperature was fixed to $k\Te = 70$ keV. The solid curves
in panel (d) show the systematic effect for $k\Te = 30$ keV 
(upper) and $k\Te = 150$ keV (lower).
\label{multi}}
\medskip

To illustrate the difference between the  Comptonization  and the power law
models,  we show the data for two  observations  together  with  the  model
spectra in  Figure~\ref{spectra}.  (A gaussian line at 6.4 keV is added
to both models.)  We note that the power law  model strongly  overestimates
the  low  energy  part  of  the  hard  component,  and  therefore  strongly
underestimates  the amplitude of the blackbody component  (dotted  curves).  
For example, the amplitude of the blackbody in the power law  fit is 
by a factor of
four too low in the upper panels (marked  Lull).  Another  strong effect is
the  difference  in the $\NH$  determined  with the two  models.  Since the
power law continues to low energies without a cutoff, the best fit $\NH$ is
$(6-7)\times 10^{22}\cminvsq$,  while for the Comptonization model $\NH\sim
2.5\times 10^{22}\cminvsq$.  The  power law  model  overestimates  the
total, corrected for absorption,  luminosity if one extrapolates  the model
spectrum   to  low   energies and gives also a poorer fit to the 
hard Lull-spectrum  (compare   left  and   right   panels   in
Fig.~\ref{spectra}). 


\bigskip
\centerline{\epsfxsize=8.5cm \epsfbox{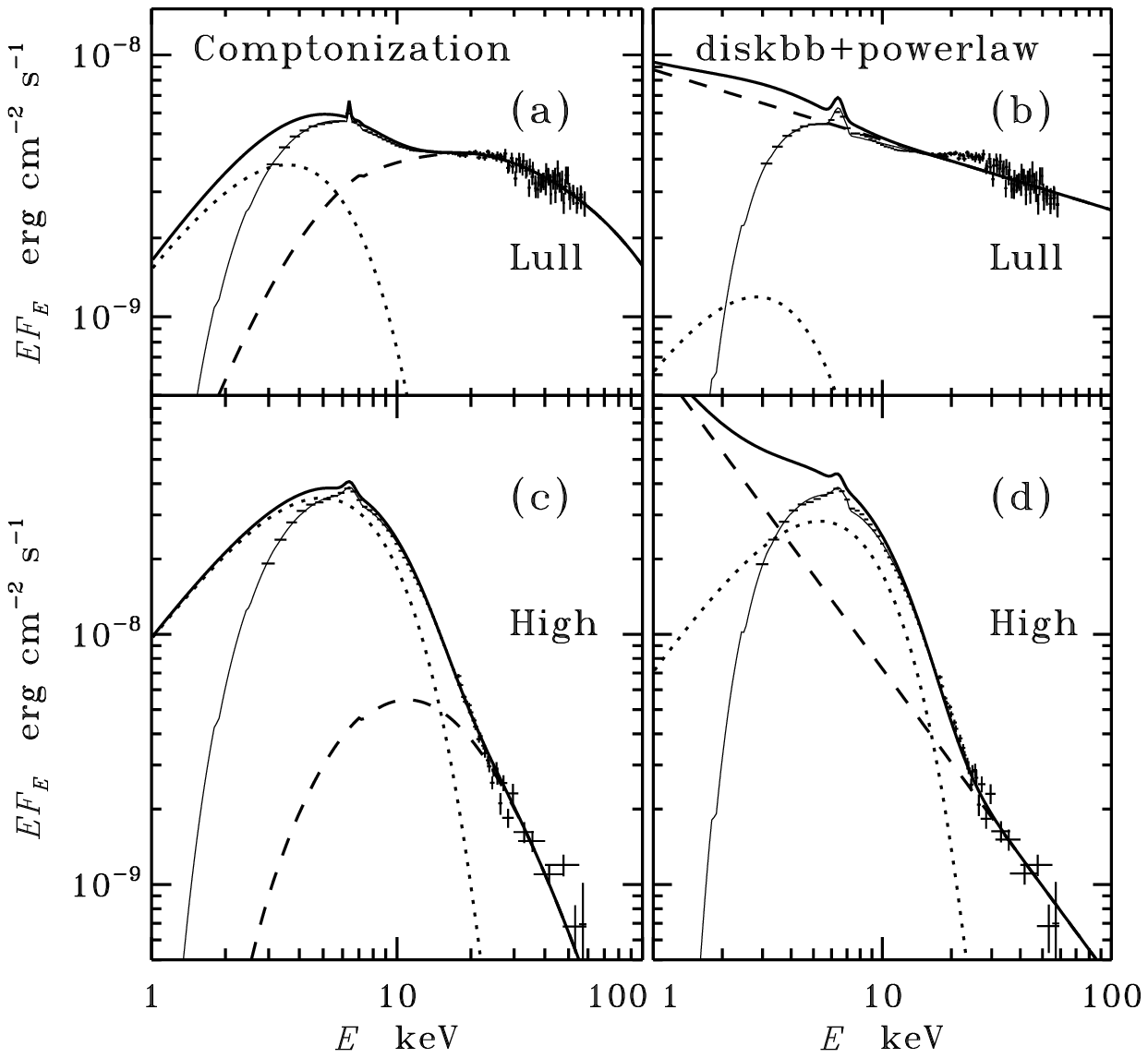}}
\figcaption{
Examples of the observed spectra, $EF_E$, of GRS 1915+105 
 in the 'Lull-phase' (1997 February 22)  and 
 in the 'High-phase' (1997 June 18).  The \RXTE\ data are shown by crosses 
 (the HEXTE data are rescaled to the PCA data). 
 The best-fit Comptonization model spectra are shown by thin solid 
 curves in  panels (a) and (c). The disk blackbody, 
 the Comptonized (plus Compton reflection) component, 
 and the total spectrum are shown by dotted, dashed, and thick solid curves,
 respectively (in all the  absorption is removed). 
 The best-fit diskbb + powerlaw model spectra are shown in panels (b) and (d)
 by thin solid curves. Here the disk blackbody, the power law, and 
 the total spectrum are shown by dotted, dashed, and thick solid curves,
 respectively (with absorption removed). 
 See text for details.
\label{spectra}}
\medskip

\section{Discussion and Conclusions}
\label{sec:disc}

A strong  correlation  between  the  ratio  $\Rin/\Rc$  and the  luminosity
$\Ldisk$  means  that  the  coronal  size  $\Rc$  decreases  when  $\Ldisk$
increases  (since  $\Rin\approx$  const).  Physically  this means  that the
range of the disk surface where energy is dissipated in the optically  thin
phase (corona)  shrinks.  When spectra are relatively hard, the geometry of
the system is similar  to a  disk-corona.  For large  $\Ldisk$  and  softer
spectra,  the  Comptonizing  plasma  does  not  cover  the  disk.  One  can
speculate that in this case hard X-rays are produced in the innermost  part
of the jet-like structure within the disk inner radius.

We cannot be  certain,  however,  that the  change  in the  geometry  (i.e.
$\Rin/\Rc$)  obtained  from the  fitting is real.  The  problem  is that we
assumed  a  constant  $\Te$  (since  the  \RXTE\  data do not  allow  us to
determine  $\Te$  unambiguously).  If one fixes $\Te$ at another value, the
result does not change  qualitatively  (see  Fig.~\ref{multi}d  where a
systematic  effect on $\Rin/\Rc$ is shown by the solid curves for $k\Te=30$
keV and 150  keV).  However,  if $\Te$ is  allowed  to vary  the  situation
changes.  

Let us consider a case of large $\Ldisk$ when $\Rin/\Rc  \approx
1$ and  $\taut \approx 0.25$ (for $k\Te=70$  keV).  
In this case  only about 3\% of soft
disk photons are  scattered in the corona.  The same  fraction is scattered
also if $\Rin/\Rc =0.5$ (here about $1/4$ of soft photons are produced within
the  corona)  and  $\taut\approx  0.03$.  Then  the  normalization  of  the
Comptonized spectrum relative to the blackbody emission is about the same.
The spectral shape does not change much since for small optical  depths and
mildly  relativistic  temperatures the  contribution of multiply  scattered
photons is small.  Therefore, the electron  temperature  has to increase to
only about 90 keV in order to produce a similar Comptonized spectrum in the
3--70 keV energy range.  From this  analysis we can only  conclude  that the
scattered fraction  decreases when $\Ldisk$  increases.  If the geometry of
the system does not change, i.e.  $\Rin/\Rc={\rm  constant}$,  $\taut$  should
decrease much more than is shown in Fig.~\ref{multi}b.  One should note
that this  discussion  is  meaningful  only if the spectra  are indeed
produced by thermal Comptonization.

Our choice of the disk  temperature profile  ($\Tbb$ is constant inside the
corona) has also some  (small)  effect on  $\Rin/\Rc$,  while the effect is
completely  negligible for other fitting  parameters.  For the most extreme
hard spectra (marked Lull) when  $\Rin/\Rc=0.5$, the number of seed photons
is doubled if the peak temperature is reached at $\Rin$ (instead of $\Rc$),
so that  $\taut$  should be  smaller  by a factor of 2 in order to  produce
similar spectra.

The mass  accretion  rate in GRS  1915+105  is high  amounting  to  $(0.1 -
1.5)\times  \dot{M}_{\rm  Edd}$ (for a  $10M_{\odot}$  black  hole),  where
$\dot{M}_{\rm  Edd} = 10\Ledd / c^2 = 1.39 \times  10^{18}  M/M_{\odot}$  g
s$^{-1}$  corresponds  to a  radiative  efficiency  of 0.1.  The inner disk
radius  lies in a rather  narrow  range of 25--40  km and the soft and hard
state radii do not differ considerably.

One should be  cautious  with the  absolute  $\Rin$-values  derived,  since
relativistic  and color  corrections  can be large.  Using the  results  by
Zhang et al.  (1997) at inclination  $70^{\grad}$,  one obtains the factors
1.23 and  1.07  for a  non-rotating  and an  extreme  prograde  Kerr  hole,
respectively,  by  which  the  radii  in  Fig.~\ref{multi}e  should  be
multiplied  to  obtain  the  innermost  radius  of the  disk.  After  these
corrections  the mass of GRS 1915+105  should be smaller than  $4M_{\odot}$
and $22M_{\odot}$  for a non-rotating and an extreme prograde Kerr hole, if
the average radius derived (30 km in  Fig.~\ref{multi}e)  equals  the
last marginally  stable orbit ($3\Rg$ and $0.5\Rg$ for a  non-rotating  and Kerr
hole, respectively).

In conclusion,  the \RXTE\ data of GRS  1915+105 in a broad  luminosity  range
$(2-30)\times  10^{38}\lum$  can be modeled  with a thermal  Comptonization
model.  We find a strong  correlation  between the spectral slope, the disk
temperature,  and the  disk  luminosity  $\Ldisk$.  The  hard  Comptonized
luminosity decreases somewhat when the total luminosity  increases (meaning
that the  fraction of total  power  dissipated  in an  optically  thin  plasma
decreases substantially with luminosity).  The scattered  fraction of seed
soft photons and the optical depth of the Comptonizing plasma decrease with
$\Ldisk$.  There are indications  that the size of the corona $\Rc$ shrinks
at large luminosities, but this can be mimicked by corresponding changes in
the optical depth.  Detailed  modeling of the broadband  spectra (at least
up to 300 keV) is needed to discriminate between those possibilities.  The
biggest question is, however, whether a pure {\it thermal}  Comptonization model
is applicable  to  GRS~1915+105.  

Finally, we would like to point out
that  spectral  fits using {\sc  xspec}  diskbb + power law  model 
underestimate   the  amplitude  of  the  blackbody   component 
and therefore the  corresponding  size of the emitting  region. 
 This  model  also 
overestimates  the absorption  column density and the luminosity  corrected
for absorption.

\acknowledgements

This  research has been  supported  by the Academy of Finland  (OV) and the
Swedish  Natural  Science  Research  Council and the Anna-Greta  and Holger
Crafoord  Fund (JP).  JN  acknowledges  the grant from ESA.  We thank Diana
Hannikainen  for reading  the  manuscript  and the  anonymous  referee  for
valuable criticism and comments.

\end{document}